\documentclass[12pt]{iopart}
\usepackage{epsfig,float,afterpage}
\newcommand{\be}{\begin{equation}}
\newcommand{\ee}{\end{equation}}
\newcommand{\bea}{\begin{eqnarray}}
\newcommand{\eea}{\end{eqnarray}}
\newcommand{\bra}{\langle}
\newcommand{\ket}{\rangle}
\newcommand{\ssz}{\scriptsize}
\newcounter{saveeqn}
\newcommand{\alpheqn}{\setcounter{saveeqn}{\value{equation}}%
\setcounter{equation}{0}%
\addtocounter{saveeqn}{1}%
\renewcommand{\theequation}{\mbox{\arabic{section}.\arabic{saveeqn}\alph{equation}}}%
}
\newcommand{\reseteqn}{\setcounter{equation}{\value{saveeqn}}%
\renewcommand{\theequation}{\arabic{section}.\arabic{equation}}}
\newcommand{\seceq}{\setcounter{equation}{0}}
\begin{document}
\jl{3}
\title{Magnetic impurities in gapless Fermi systems: perturbation theory}
\author{Matthew T Glossop and David E Logan}
\address{Oxford University, Physical and Theoretical Chemistry Laboratory, South Parks Road, Oxford OX1 3QZ, UK}
\begin{abstract}
We consider a symmetric Anderson impurity model with a soft-gap hybridization vanishing at the Fermi level, $\Delta_{\mbox{\ssz{I}}}\propto |\omega|^r$ with $r>0$.  Three facets of the problem are examined.  First the non-interacting limit, which despite its simplicity contains much physics relevant to the $U>0$ case: it exhibits both strong coupling (SC) states (for $r<1$) and local moment states (for $r>1$), with characteristic signatures in both spectral properties and thermodynamic functions.  Second, we establish general conditions upon the interaction self-energy for the occurence of a SC state for $U>0$. This leads to a pinning theorem, whereby the modified spectral function $A(\omega)=|\omega|^rD(\omega)$ is pinned at the Fermi level $\omega = 0$ for any $U$ where a SC state obtains; it generalizes to arbitrary $r$ the pinning condition upon $D(\omega=0)$ familiar in the normal $r=0$ Anderson model.  Finally, we consider explicitly spectral functions at the simplest level: second order perturbation theory in $U$, which we conclude is applicable for $r<\frac{1}{2}$ and $r>1$ but not for $\frac{1}{2}<r<1$.  Characteristic spectral features observed in numerical renormalization group calculations are thereby recovered, for both SC and LM phases; and for the SC state the modified spectral functions are found to contain a generalized Abrikosov-Suhl resonance exhibiting a characteristic low-energy Kondo scale with increasing interaction strength. 
\end{abstract}
\maketitle
\section{Introduction}
The Kondo effect, whereby an impurity spin is quenched by coupling to the low-energy excitations of a non-interacting metallic host, has long occupied a central role in the study of magnetic impurities (see e.g. [1]).  The effect is of course normally regarded as being dependent upon a metallic host, with low-energy and hence low-temperature impurity properties controlled by the non-vanishing host density of states at the Fermi level, $\omega = 0$, and essentially independent of the details of host band structure.

But what if the host exhibits semi-metallic character, with a spectrum whose low-energy behaviour exhibits a soft-gap at the Fermi level, $\rho_{\mbox{\ssz{host}}} \propto |\omega|^r$ with $r>0$?  There are quite a number of experimental candidates for such behaviour, ranging from semiconductors whose valence and conduction bands touch at the Fermi level [2], through heavy Fermion superconductors [3], to various two-dimensional systems including graphite sheets [4] and quasi-one-dimensional metals described by a Luttinger model [5].

The question above was first posed by Withoff and Fradkin [6], who studied the soft-gap Kondo model using a combination of `poor man's' scaling and a large-$N$ mean-field theory (with $N$ the impurity degeneracy).  Since then, there has been much study of both the Kondo and the corresponding Anderson impurity models, in particular via scaling [6,7], large-$N$ expansions [6,8,9] and the numerical renormalization group (NRG) [7,10-13].  All these techniques, whether for the soft-gap Anderson or Kondo models, show the existence of two distinct types of ground state, between which in general a nontrivial zero-temperature phase transition occurs at a finite value of the host-impurity coupling (or, equivalently in the Anderson model, at a finite value of the impurity on-site interaction, $U$): a weak coupling or local moment (LM) state in which the impurity spin remains unquenched; and a strong coupling (SC) state in which a Kondo effect is manifest, and whose properties---in particular for the so-called symmetric strong coupling state considered here---have been argued to represent a natural generalization of Fermi liquid physics (see especially [13]).

NRG studies in particular have devoted considerable attention to the spin-$\frac{1}{2}$ ($N=2$) particle-hole symmetric case, including thermodynamic [10-13] properties and, for the Anderson model, impurity spectral functions [12].  It is the symmetric spin-$\frac{1}{2}$ soft-gap Anderson model we consider here, with aims that are modest, and threefold.  First, to consider briefly the non-interacting limit, $U=0$; second to establish rather general conditions upon the existence of a SC state at finite-$U$; and finally, to examine the problem explicitly at the simplest possible level---straight second-order perturbation theory in $U$.

There are two reasons for considering the non-interacting limit, simple though it is.  First, and in contrast to the `normal' Anderson model ($r=0$), its behaviour is nontrivial and exemplifies much physics relevant to the interacting problem.  An understanding of the impurity single-particle spectrum, $D_0(\omega)$, is sufficient to study the non-interacting limit, since both `excess' thermodynamic properties induced by addition of the impurity, and local properties such as the impurity susceptibility, follow from a knowledge of it.  As shown in section 3, both LM and SC states arise in the $U = 0$ limit, with characteristic and distinct signatures in both spectral and thermodynamic functions, as indeed known in part from the work of ref. [13].  LM states alone are found to occur for $r>1$ and SC states for $r<1$, whereas finite-$U$ NRG results [12,13] show that LM states alone occur for $r>\frac{1}{2}$---an important contrast whose implications are considered in section 5.  The second reason for considering the non-interacting limit is prosaic: a knowledge of it underpins finite-order perturbation theory in $U$, as considered in section 5.

In section 4, focusing on the finite-$U$ impurity spectral function $D(\omega)$, we establish general conditions upon the interaction self-energy for the occurence of a SC state, from which follow in turn two results.  First, that the low-frequency behaviour of the single-particle spectrum is $D(\omega) \sim |\omega|^{-r}$, which is precisely the spectral signature of the SC state found in NRG studies [12].  Second, and relatedly, that interactions have no influence in renormalizing the low-$\omega$ asymptotic behaviour of $D(\omega)$.  In consequence, one obtains a conservation on $A(\omega)=|\omega|^rD(\omega)$ at the Fermi level, $\omega = 0$: for any $r$ where a SC state exists, $A(\omega=0)$ is pinned at its non-interacting value for all $U$, a result that generalizes to arbitrary $r$ the corresponding condition familiar for the $r=0$ Anderson model (see e.g.[1]).

In contrast to the $r=0$ Anderson model where the predictions of perturbation theory in $U$ about the non-interacting limit are well known (see e.g.[1]), the implications---and, indeed, general applicability---of a low-order perturbative treatment are not obvious for the soft-gap problem, and are considered in section 5 where we focus on the impurity spectral function.  For $r>1$ second-order perturbation theory is found to recover the characteristic low-$\omega$ spectral signature of the LM state found in NRG studies [12], viz $D(\omega) \sim |\omega|^r$; and the resultant single-particle spectra are investigated in some detail.  For $r< \frac{1}{2}$ we find the SC state is indeed perturbatively stable upon increasing $U$ from zero, and that the general conditions of section 4 for a SC state are satisfied at the second-order level; the resultant modified spectral functions $A(\omega)$ are also shown to bear a striking resemblance to that for the normal Anderson model, $r = 0$, exhibiting in particular the emergence with increasing $U$ of a characteristic low-energy Kondo scale.  For $\frac{1}{2}<r<1$ by contrast, we argue that finite-order perturbation theroy in $U$ about the non-interacting limit is simply inapplicable.

There is a second, `hidden' reason why we consider low-order perturbation theory: to illustrate its limitations, despite its strengths, even for $r< \frac{1}{2}$ where the SC state is perturbatively continuable from the non-interacting limit.  It is our belief that to describe analytically much of the underlying physics of the soft-gap Anderson model---and in particular to capture the transition between LM and SC (or generalized Fermi liquid) phases which render the problem of generic interest---requires, or at least invites, a new and inherently non-perturbative theoretical approach.  We will turn to one such theory in a subsequent paper [14], for which the present work is in part a forerunner.   
\seceq
\section{Background}
With the Fermi level taken as the origin of energy, the Hamiltonian for the spin-$\frac{1}{2}$ Anderson model is given in standard notation by
\alpheqn
\begin{eqnarray}
\hat{H}& = \hat{H}_{\mbox{\ssz{host}}}+\hat{H}_{\mbox{\ssz{impurity}}}+\hat{H}_{\mbox{\ssz{hybridization}}}\\
& = \sum_{\bi{k}, \sigma}\epsilon_{\bi{k}}\hat{n}_{\bi{k} \sigma}+\sum_{\sigma}(\epsilon_{i}+\frac{U}{2}\hat{n}_{i-\sigma})\hat{n}_{i\sigma}+\sum_{\bi{k},\sigma}V_{i\bi{k}}(c^{\dagger}_{i\sigma}c_{\bi{k}\sigma}+c^{\dagger}_{\bi{k}\sigma}c_{i\sigma})
\end{eqnarray}
\reseteqn
with $\epsilon_{\bi{k}}$ the host dispersion, $V_{i\bi{k}}$ the hybridization and $\epsilon_{i}$ the impurity level; for the symmetric case considered here, $\epsilon_{i} = - \frac{U}{2}$ with $U$ the on-site interaction.

We consider the zero-temperature single-particle impurity Green function, $G(t) = -\mbox{i}\bra T\{c_{i\sigma}(t)c^{\dagger}_{i\sigma}\}\ket$, with $G(\omega)$ expressible as
\be
G(\omega) = \left[\omega + \mbox{i}\eta \mbox{sgn}(\omega)-\Delta(\omega)-\Sigma(\omega)\right]^{-1}
\ee
where the limit $\eta \rightarrow 0+$ is henceforth understood.  Here $\Delta(\omega)$ is the hybridization function given by
\alpheqn
\bea
\Delta(\omega)&=\sum_{\bi{k}}\frac{|V_{i\bi{k}}|^{2}}{\omega-\epsilon_{\bi{k}}+\mbox{i}\eta \mbox{sgn}(\omega)}\\
& = \Delta_{\mbox{\ssz{R}}}(\omega)-\mbox{isgn}(\omega)\Delta_{\mbox{\ssz{I}}}(\omega)
\eea
\reseteqn
and we consider throughout a symmetric hybridization
\be
\Delta(\omega)=-\Delta(-\omega)
\ee
(whose particular form is specified in \S2.1).  From particle-hole symmetry the Fermi level remains fixed at $\omega = 0\  \forall\  U\ge0$, whence the impurity charge $n_{i} = \sum_{\sigma}\bra\hat{n}_{i\sigma}\ket = 1 \ \forall \ U$; and the interaction self-energy $\Sigma(\omega)$ is defined to exclude the trivial Hartree contribution of $(U/2)n_{i}=U/2$, which cancels $\epsilon_{i} = -U/2$.  $\Sigma(\omega)$ may likewise be decomposed as

\be
\Sigma(\omega)=\Sigma_{\mbox{\ssz{R}}}(\omega)-\mbox{isgn}(\omega)\Sigma_{\mbox{\ssz{I}}}(\omega)
\ee
and the real parts of $\Sigma$ or $\Delta$ follow directly from the Hilbert transform

\be
F_{\mbox{\ssz{R}}}(\omega)=\int_{-\infty}^{\infty}\frac{\rmd\omega_{1}}{\pi}F_{\mbox{\ssz{I}}}(\omega_{1})\mbox{P}\left (\frac{1}{\omega-\omega_{1}}\right)
\label{ht} 
\ee
with F = $\Sigma$ or $\Delta$ as appropriate.

The single-particle impurity spectrum $D(\omega)=-\frac{1}{\pi}\mbox{sgn}(\omega)\mbox{Im}G(\omega)$ follows directly as

\be
D(\omega) = D^{\mbox{\ssz{p}}}(\omega) +  D^{\mbox{\ssz{b}}}(\omega)
\label{pb}
\ee
where
\alpheqn
\be
D^{\mbox{\ssz{b}}}(\omega) = \frac{\left[\Delta_{\mbox{\ssz{I}}}(\omega)+\Sigma_{\mbox{\ssz{I}}}(\omega)\right]\pi^{-1}}{\left[\omega-\Delta_{\mbox{\ssz{R}}}(\omega)-\Sigma_{\mbox{\ssz{R}}}(\omega)\right]^{2}+\left[\eta+\Delta_{\mbox{\ssz{I}}}(\omega)+\Sigma_{\mbox{\ssz{I}}}(\omega)\right]^{2}}
\ee
and
\be
D^{\mbox{\ssz{p}}}(\omega) = \frac{\eta \pi^{-1}}{\left[\omega-\Delta_{\mbox{\ssz{R}}}(\omega)-\Sigma_{\mbox{\ssz{R}}}(\omega)\right]^{2}+\left[\eta+\Delta_{\mbox{\ssz{I}}}(\omega)+\Sigma_{\mbox{\ssz{I}}}(\omega)\right]^{2}}.
\ee
\reseteqn
Here $D^{\mbox{\ssz{b}}}(\omega)$ refers to continuum (or `band') excitations, while  $D^{\mbox{\ssz{p}}}(\omega)$ allows for the possibility of discrete states reflected in pole contributions.

The change in the density of states of the system due to addition of the impurity, $\Delta\rho(\omega)$, will also prove central to the subsequent analysis.  It is given by
\be
\Delta\rho(\omega)=D(\omega)\left[1-\frac{\partial\Delta_{\mbox{\ssz{R}}}(\omega)}{\partial\omega}\right]-\frac{1}{\pi}\mbox{Re}G(\omega)\frac{\partial\Delta_{\mbox{\ssz{I}}}(\omega)}{\partial\omega}
\label{delrho}
\ee
and is calculable directly once the impurity $D(\omega)$ is known.  Note that equation (\ref{delrho}), while commonly derived explicitly for the non-interacting case (see eg [1]), is readily shown to hold generally for all $U$.  From equation (\ref{pb}),  $\Delta\rho(\omega)$ likewise separates into `band' and `pole' contributions,
\alpheqn
\bea
\Delta\rho(\omega)=\Delta\rho^{\mbox{\ssz{p}}}(\omega)+\Delta\rho^{\mbox{\ssz{b}}}(\omega)
\eea
where in particular
\bea
\Delta\rho^{\mbox{\ssz{p}}}(\omega)= D^{\mbox{\ssz{p}}}(\omega)\left[1-\frac{\partial\Delta_{\mbox{\ssz{R}}}(\omega)}{\partial\omega}\right].
\eea
\reseteqn
\subsection{Hybridization function}
The hybridization function we consider, $\Delta_{\mbox{\ssz{I}}}(\omega)\left(=\Delta_{\mbox{\ssz{I}}}(-\omega)\right)$, is given by
\be
\Delta_{\mbox{\ssz{I}}}(\omega) = \left\{\begin{array}{ccc}
\Delta_{0}\left(|\omega|-\frac{\delta}{2}\right)^{r} &:&\frac{\delta}{2}<|\omega|<D+\frac{\delta}{2}\\
0 &:& \mbox{otherwise}
\end{array}
\right.
\label{gap}
\ee
with $r>0$.  While a pure power-law hybridization is naturally not realistic on arbitrary energy scales, it captures in the simplest fashion the requisite low-$\omega$ behaviour; moreover, as for the $r=0$ Anderson model, one expects impurity properties to be controlled primarily by the low-$\omega$ behaviour and to be largely independent of the details of host band structure.  In general the hybridization thus contains a gap of magnitude $\delta$, in which lies the Fermi level $\omega = 0$; the `normal' flat-band Anderson model is recovered as a special case of equation (\ref{gap}) with $r = 0 = \delta$.  There are four energy scales in the problem, namely $\delta$, $\Delta_{0}^{\frac{1}{1-r}}$, $D$ (the bandwidth) and $U$; we choose to rescale in terms of $\Delta_{0}^{\frac{1}{1-r}}$, defining for later purposes
\be
\tilde{\omega} = \frac{\omega}{\Delta_{0}^{\frac{1}{1-r}}},\,\,\,\,\,\tilde{D} = \frac{D}{\Delta_{0}^{\frac{1}{1-r}}},\,\,\,\,\,\tilde{U} = \frac{U}{\Delta_{0}^{\frac{1}{1-r}}}.
\ee

From the Hilbert transform equation (\ref{ht}), $\Delta_{\mbox{\ssz{R}}}(\omega) = -\Delta_{\mbox{\ssz{R}}}(-\omega)$ is given by
\be
\Delta_{\mbox{\ssz{R}}}(\omega)=\frac{2\omega\Delta_{0}}{\pi}\int_{0}^{D} \rmd\omega_{1}\ \frac{\omega_{1}^r}{\omega^{2}-\left(\omega_{1}+\frac{\delta}{2}\right)^{2}}
\label{delr}
\ee
where a principal value is henceforth understood.  Notice from this that the wide-band limit $D\rightarrow\infty$, as commonly employed for the normal Anderson model $r = 0 = \delta$, can be taken only for $r<1$.

For the gapped case, $\delta > 0$, we shall need solely the behaviour of $\Delta_{\mbox{\ssz{R}}}(\omega)$ for frequencies $|\omega|\ll\frac{\delta}{2}$ inside the gap.  This is given from equation (\ref{delr}) by
\alpheqn
\be
\Delta_{\mbox{\ssz{R}}}(\omega) = -\omega\frac{2\Delta_{0}}{\pi}\left[\frac{\delta}{2}\right]^{r-1}B(\delta;r)+\Or
\left[\left(\frac{2\omega}{\delta}\right)^{3}\right]
\ee
where $B(\delta;r)\geq 0$ is given by
\be
B(\delta;r)= \int_{0}^{2D/\delta}\rmd z \frac{z^r}{\left(1+z\right)^2}.
\ee
\reseteqn

Our primary focus in sections 4 and 5 will be the `soft gap' case, $\delta = 0$, where $\Delta_{\mbox{\ssz{I}}}(\omega)=\Delta_{0}|\omega|^r$.  For this case, rescaling of equation (2.13) yields
\be
\Delta_{\mbox{\ssz{R}}}(\omega)= \mbox{sgn}(\omega)\Delta_{0}|\omega|^r\frac{2}{\pi}\int_{0}^{D/|\omega|}\rmd y\  \frac{y^r}{\left(1-y^2\right)}
\label{215}
\ee
which thus obeys the differential equation
\be
\frac{\partial\Delta_{\mbox{\ssz{R}}}(\omega)}{\partial\omega}=\frac{r}{\omega}\Delta_{\mbox{\ssz{R}}}(\omega)+\frac{2\Delta_{0}D^{r-1}}{\pi}\left[1-\left(\frac{\omega}{D}\right)^2\right]^{-1}.
\ee
Direct evaluation of equation (\ref{215}) yields, for $|\omega|<D$,
\alpheqn
\bea
\fl\ \ \ \ \ \ \ \ \ \Delta_{\mbox{\ssz{R}}}(\omega)&= -\mbox{sgn}(\omega)\Delta_{0}\left\{\tan \left(\mbox{$\frac{\pi r}{2}$}\right)|\omega|^r+\frac{D^r}{\pi r}\biggr[F(1,-r;1-r;|\omega|/D)\biggr.\right.  \nonumber\\
&\ \ \ \ \ \ \ \ \ \ \ \ \ \ \  -\biggr.\biggr. F(1,-r;1-r;-|\omega|/D)\biggr]\biggr\}\\
& = -\mbox{sgn}(\omega)\Delta_{0}\left\{\mbox{tan}\left(\mbox{$\frac{\pi r}{2}$}\right)|\omega|^r+\frac{2D^r}{\pi(r-1)}\frac{|\omega|}{D}+\Or\left[\left(\frac{|\omega|}{D}\right)^3\right]\right\}
\label{hgb}
\eea
\reseteqn
with $F(\alpha,\beta;\gamma;z)$ a Gauss hypergeometric function.  Equation (2.17) encompasses the results given in [13] and applies for any $r \geq 0$, including $r=1$ for which the limit as $r\rightarrow 1$ of equation (\ref{hgb}) gives
\be
\Delta_{\mbox{\ssz{R}}}(\omega)=\frac{2\Delta_0}{\pi}\mbox{sgn}(\omega)|\omega|\ \ln\left(\mbox{$\frac{|\omega|}{D}$}\right) + \Or\left[\left(\frac{|\omega|}{D}\right)^3\right]\,\,\,\,:\,\,\,\,r = 1
\ee
showing the characteristic logarithmic behaviour that, as discussed in section 3, is indicative of the marginal nature of $ r = 1$ for $U = 0$.
\seceq
\section{Non-interacting limit}
As mentioned in section 1, the non-interacting problem is surprisingly rich: it contains already much underlying physics relevant to the interacting case and, for the soft-gap case in particular, gives rise to both SC and LM states as now considered.

The impurity spectrum $D_0(\omega)$ (with `0' referring to $U = 0$) is the primary quantity, since the excess density of states $\Delta\rho_{0}(\omega)$ follows from it via equation (\ref{delrho}).  The low-$\omega$ behaviour of the latter determines in turn the low temperature behaviour of the change in thermodynamic properties due to addition of the impurity; the `excess' total uniform spin susceptibility, specific heat and entropy being given trivially for non-interacting electrons by
\alpheqn
\be
\chi^{0}_{\mbox{\ssz{imp}}}(T)=\frac{(g\mu_B)^2}{2T}\int_{-\infty}^{\infty}\rmd\omega\ \Delta\rho_0(\omega)f(\omega)\left(1-f(\omega)\right)\\
\ee
\be
C^0_{\mbox{\ssz{imp}}}(T) = \frac{2}{T^2}\int_{-\infty}^{\infty}\rmd\omega\  \omega^2\Delta\rho_0(\omega)f(\omega)\left(1-f(\omega)\right)\\
\ee
\be
\fl \ \ \ \ \ S^0_{\mbox{\ssz{imp}}}(T)= -2\int_{-\infty}^{\infty}\rmd\omega\ \Delta\rho_0(\omega)\left[f(\omega)\ln f(\omega)+\left(1-f(\omega)\right)\ln \left(1-f(\omega)\right)\right]
\ee
\reseteqn
where $f(\omega)$ is the Fermi function (with chemical potential $\mu = 0$ for all $T$ due to particle-hole symmetry), and $k_B \equiv 1$ has been taken.

The band/continuum part of $\Delta\rho_0(\omega)$ is given from equations (2.9,10) by
\be
\Delta\rho_0^{\mbox{\ssz{b}}}(\omega)= D_0^{\mbox{\ssz{b}}}(\omega)\left[1-\frac{\partial\Delta_{\mbox{\ssz{R}}}(\omega)}{\partial\omega}\right]-\frac{1}{\pi}\mbox{Re}G_0(\omega)\frac{\partial\Delta_{\mbox{\ssz{I}}}(\omega)}{\partial\omega}
\ee
with $D_0^{\mbox{\ssz{b}}}(\omega)$ from Eq(2.8a).  For the gapless case $\delta = 0$, a straightforward calculation using Eq(2.16) for $\partial\Delta_{\mbox{\ssz{R}}}/\partial\omega$ gives a simple relation between $\Delta\rho_0^{\mbox{\ssz{b}}}(\omega)$ and $D_0^{\mbox{\ssz{b}}}(\omega)$ for $|\omega| < D$: 
\alpheqn
\be
\Delta\rho_0^{\mbox{\ssz{b}}}(\omega) = D_0^{\mbox{\ssz{b}}}(\omega)\frac{(1-r)}{q(\omega)}\ \ \ \ :\ \ \ \ \delta = 0
\ee
where
\be
q^{-1}(\omega)=\left[1+\frac{2\Delta_0D^{r-1}}{\pi(r-1)}\frac{1}{1-\left(\frac{\omega}{D}\right)^2}\right].
\ee
\reseteqn
The pole contributions, $D_0^{\mbox{\ssz{p}}}(\omega)$ and $\Delta\rho_{0}^{\mbox{\ssz{p}}}(\omega)$, are considered below.  Here we simply note that for finite bandwidth, $D$, there are always such contributions outside the band, $|\omega|>D$.  These however are of no importance to the problem, and are not considered explicitly in what follows where $|\omega|<D$ is implicit.

\subsection{Gapped case: LM state}
We consider first the case of an insulating host, with $\delta>0$.  Since $\Delta_{\mbox{\ssz{I}}}(\omega)=0$ for $|\omega|<\delta/2$ inside the gap, it follows from equation (2.8b) (with $\Sigma = 0$) that
\be
D_{0}^{\mbox{\ssz{p}}}(\omega) = q\delta(\omega)
\ee
with poleweight $q$ given by
\alpheqn
\bea
q^{-1} =& \left[1-\left(\frac{\partial\Delta_{\mbox{\ssz{R}}}(\omega)}{\partial\omega}\right)_{\omega=0}\right]\\
       =& 1+\frac{2\Delta_0}{\pi}\left[\frac{\delta}{2}\right]^{r-1}B(\delta;r)
\eea
\reseteqn
where equation (2.14) is used.  From equation (2.8b) the band contribution $D_0^{\mbox{\ssz{b}}}(\omega)$ naturally vanishes inside the gap, and behaves as $D_0^{\mbox{\ssz{b}}}(\omega)\propto \Delta_{\mbox{\ssz{I}}}(\omega)\sim (|\omega|-\delta/2)^r$ close to the gap edges.  $D_0(\omega)$ is thus dominated by the discrete state at the Fermi level, $\omega = 0$, as too is $\Delta\rho_0(\omega)$: from equations (2.10b) and (3.4,5),
\be
\Delta\rho_0^{\mbox{\ssz{p}}}=\delta(\omega)
\ee
whose poleweight of unity reflects a `whole' single extra state at the Fermi level induced by addition of the impurity.

These features---a whole excess state at the Fermi level, with non-vanishing weight on the impurity---are the hallmark of the LM state for $U = 0$.  They naturally control the low-$T$ excess thermodynamic functions, given from equations (3.1) by 
\alpheqn
\be
\chi^{0}_{\mbox{\ssz{imp}}}(T)=\frac{(g\mu_B)^2}{8kT}=\mbox{$\frac{1}{2}$}\chi_{\mbox{\tiny{Curie}}}(T)
\ee
\be
S^{0}_{\mbox{\ssz{imp}}}(0)=\ln4
\ee
\reseteqn
with corrections that are thermally activated (as too is $C^{0}_{\mbox{\ssz{imp}}}(T))$.  The residual entropy of $\ln4$ is physically obvious: the excess Fermi level state has four occupancies---empty, $\uparrow$-spin or $\downarrow$-spin occupied and doubly occupied---with equal a priori probabilities for $U = 0$.  This is also why $\chi^{0}_{\mbox{\ssz{imp}}}(T)$ is {\it half} a Curie law: only half of the four occupancies, the singly occupied states, are paramagnetically active.  This situation will of course change immediately for any $U>0$ where doubly occupied (and hence empty) spin configurations are suppressed, producing instead $\chi_{\mbox{\ssz{imp}}}(T)=\chi_{\mbox{\tiny{Curie}}}(T)$ as $T\rightarrow 0$, and $S_{\mbox{\ssz{imp}}}(0)=\ln 2$.

\restylefloat{figure}
\afterpage{\begin{figure}[H] 
\centering 
\epsfig{file =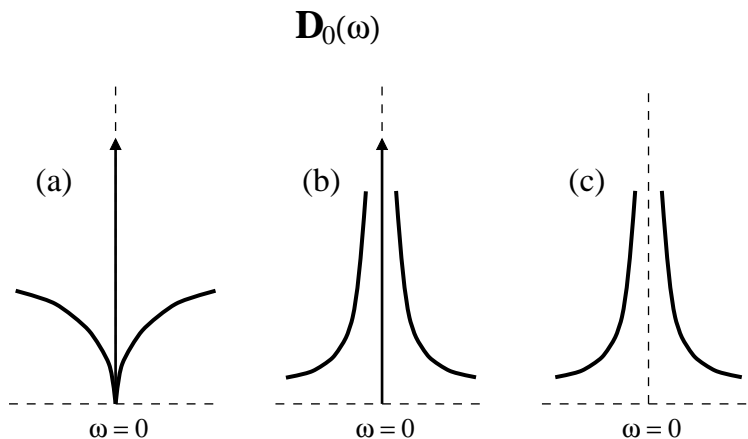,width=10cm} 
\caption{Schematic of non-interacting impurity spectra $D_0(\omega)$ for (a) LM state, $r>2$; (b) LM state, $1<r<2$; (c) SC state, $r<1$.}
\end{figure}}

The gapless case, $\delta = 0$, is considered in sections 3.2,3, but first we ask what happens as $\delta \rightarrow 0$.  From equation (2.14),
\alpheqn
\be
B(\delta;r)=\frac{\pi r}{\sin(\pi r)}+\frac{1}{(r-1)}\ \frac{x^{1-r}}{(1+x)}\ F(1,-r;2-r;-x)
\ee
with
\be
x = \frac{\delta}{2D}. 
\ee
\reseteqn
From equation (3.5b), the behaviour of the poleweight $q$ for $\delta/2D \ll 1$ can thus be obtained.  For $r>1$ one finds
\be
q^{-1}=\left[1+\frac{2\Delta_0D^{r-1}}{\pi(r-1)}\right]+\Or\left(\delta^{r-1};\delta\right) \ \ \ \ \ : r>1
\ee
which remains finite as $\delta \rightarrow 0$ (where the resultant $q = q(\omega = 0)$, from equation (3.3b)), showing the persistence of the LM state for $r > 1$ when $\delta = 0$.  For $r < 1$ by contrast,
\be
q^{-1} = 1+\frac{2r}{\sin(\pi r)}\Delta_0D^{r-1}\left[\frac{\delta}{2D}\right]^{r-1}+\Or(1)\ \ \ \ : \ \ \ \ r<1
\ee
which diverges as $\delta \rightarrow 0$ (as too does the marginal case of $r = 1$, where \newline $q^{-1}\sim 1/\ln(2D/\delta)$).  For $r<1$, the LM state does not therefore survive closure of the gap.

\subsection{Gapless case: $r>1$, LM state}
For $r>1$, the LM state indeed persists with the gap shut, $\delta = 0$: since $\Delta_{\mbox{\ssz{I}}}(\omega)\sim|\omega|^r$ decays to zero as $\omega \rightarrow 0$ more rapidly than $\Delta_{\mbox{\ssz{R}}}(\omega)\sim \Or(|\omega|)$ (see equation (2.17b)), equation (2.8b) yields directly $D_0^{\mbox{\ssz{p}}} = q\delta(\omega)$ with $q$ given by equation (3.5a) and hence equation (3.9) with $\delta = 0$. Likewise, from equation (2.10b), $\Delta\rho_0^{\mbox{\ssz{p}}}(\omega) = \delta(\omega)$ contains a `whole' extra state.

In contrast to the gapped case however, $D_0^{\mbox{\ssz{b}}}(\omega)$ now extends down to $\omega = 0$.  From equations (2.8a) and (2.17b) one obtains
\be
D_0^{\mbox{\ssz{b}}}(\omega)=\frac{\Delta_0q^2}{\pi}|\omega|^{r-2}+\Or(|\omega|^{2r-3})
\ee
diverging as $\omega \rightarrow 0$ for $1<r<2$ and vanishing for $r>2$, as found in [13]. The full $D_0(\omega)$ consists of course of both pole and band contributions; it is illustrated schematically in figure 1.  From equations (3.3) and (3.11) the low-$\omega$ behaviour of $\Delta\rho_0^{\mbox{\ssz{b}}}(\omega)$ is
\be
\Delta\rho_0^{\mbox{\ssz{b}}}(\omega)=\frac{\Delta_0q}{\pi}(1-r)|\omega|^{r-2}+\Or\left(|\omega|^{2r-3};|\omega|^r\right).
\ee
The excess thermodynamic properties follow directly via equation (3.1): $S^0_{\mbox{\ssz{imp}}}(0)=\ln4$ again, while
\alpheqn
\be
\chi^0_{\mbox{\ssz{imp}}}(T)=\mbox{$\frac{1}{2}$}\chi_{\mbox{\tiny{Curie}}}(T)+c'T^{r-2}+\Or(T^{2r-3};T^r)
\ee
\be
C^0_{\mbox{\ssz{imp}}}(T)=d'T^{r-1}+\Or(T^{2(r-1)};T^{1+r})
\ee
\reseteqn
where $c'$ and $d'$ are negative constants (whose sign reflects the low-$\omega$ `depletion' of $\Delta\rho_0^{\mbox{\ssz{b}}}(\omega)$---the coefficient of the leading $|\omega|^{r-2}$ term in equation (3.12) being negative for $r>1$).  Not surprisingly, and in part for the reasons discussed in section 3.1, the behaviour equation (3.13) characteristic of the $U = 0$ limit differs from that found by NRG for the $U>0$ LM regime [12]: $\chi_{\mbox{\ssz{imp}}}(T) = \chi_{\mbox{\tiny{Curie}}}(T)+c'_{3}T^{r-1}$ and $C_{\mbox{\ssz{imp}}}(T)\propto T^r$.

\subsection{Gapless case: $r<1$, SC state}
From equations (2.8a) and (2.17b), a simple calculation gives the low-$\omega$ behaviour of the continuum contribution to $D_0(\omega)$, viz
\alpheqn
\be
D_0^{\mbox{\ssz{b}}}(\omega)=\frac{|\omega|^{-r}}{\pi\Delta_0(1+\beta^2)}\left[1-\frac{\beta}{1+\beta^2}\ \frac{2}{\Delta_0q(0)}\ |\omega|^{1-r}\right]+\Or(|\omega|^{2-3r})
\ee
where
\be
\beta = \tan \left(\mbox{$\frac{\pi}{2}r$}\right).
\ee
\reseteqn
$\Delta\rho_0^{\mbox{\ssz{b}}}(\omega)$ follows directly from equation (3.3a) and has the same leading asymptotics as $D_0^{\mbox{\ssz{b}}}(\omega)$.  

Now consider $D_0^{\mbox{\ssz{p}}}(\omega)$, given from equation (2.8b) with $\Sigma = 0$.  From equation (2.17b) the leading low-$\omega$ behaviour of $\Delta_{\mbox{\ssz{R}}}(\omega)$ for $0<r<1$ may be cast as
\be
\Delta_{\mbox{\ssz{R}}}(\omega)=-\mbox{sgn}(\omega)\tan\left(\mbox{$\frac{\pi}{2}r$}\right)\Delta_{\mbox{\ssz{I}}}(\omega)\ \ \ \ \ : \ \ \ \ \omega\rightarrow 0
\ee
whence
\be
D_0^{\mbox{\ssz{p}}}(\omega)\equiv L(\Delta_{\mbox{\ssz{I}}};r)
\ee
with $L(x;\lambda)$ (used again in section 4) defined generally by
\alpheqn
\bea
L(x;\lambda) &=\frac{\eta\pi^{-1}}{\left[\tan\left(\frac{\pi}{2}\lambda\right)x\right]^2+\left[\eta+x\right]^2}\ \ \ \ :\ \ \ \ \eta\rightarrow0+\\
	     &=\frac{\lambda}{\tan\left(\frac{\pi}{2}\lambda\right)}\delta(x).
\eea
\reseteqn
Since $\Delta_{\mbox{\ssz{I}}}(\omega)=\Delta_0|\omega|^r$, it follows that
\be
D_0^{\mbox{\ssz{p}}}(\omega)=\frac{1}{\Delta_0\tan\left(\frac{\pi}{2}r\right)}|\omega|^{1-r}\delta(\omega)\ \ \ \ \ \equiv 0
\ee
i.e. there is no pole contribution to $D_0(\omega)$ itself, consistent with section (3.1) as $\delta\rightarrow 0$.  Hence $D_0(\omega)\equiv D_0^{\mbox{\ssz{b}}}(\omega)$ is given by equation (3.14), with $D_0(\omega) \sim |\omega|^{-r}$ as found in [13] (and illustrated schematically in figure 1c).

But from equation (2.10b), $\Delta\rho_0^{\mbox{\ssz{p}}}(\omega)$ is given in contrast by
\alpheqn
\bea
\Delta\rho_0^{\mbox{\ssz{p}}}(\omega)&= D_0^{\mbox{\ssz{p}}}(\omega)\left[1+\Delta_0r\tan\left(\mbox{$\frac{\pi}{2}r$}\right)|\omega|^{r-1}\right]\\
		&= r\delta(\omega)
\eea
\reseteqn
(via equation (3.18)), and thus contains a $\delta$-function contribution of weight $r$, as well known from the work of [10,13] where the result was obtained from analysis of the phase shift.  This behaviour is the characteristic of the SC state, and suggests the interpretation [13] that a fraction $r$ of a conduction electron occupies the decoupled excess state at the Fermi level, the remaining fraction $1-r$ being absorbed into the resonant continuum centred on $\omega = 0$.  And note again that, in contrast to the LM state, the excess level has no weight on the impurity itself: $D_0^{\mbox{\ssz{p}}}(\omega)\equiv0$.

Excess thermodynamic properties follow directly via equation (3.1), namely
\alpheqn
\be
\chi^0_{\mbox{\ssz{imp}}}(T)=\mbox{$\frac{r}{2}$}\chi_{\mbox{\tiny{Curie}}}(T)+cT^{-r}+\Or(T^{1-2r})
\ee
\be
C^0_{\mbox{\ssz{imp}}}(T)=dT^{1-r}+\Or(T^{2(1-r)})
\ee
\be
S^0_{\mbox{\ssz{imp}}}(T)=2r\ln2+eT^{1-r}+\Or(T^{2(1-r)})
\ee
\reseteqn
where $c$, $d$ and $e$ are positive constants.  These are worth noting for comparison to NRG results obtained for the $U>0$ SC phase, namely [10--13]
\alpheqn
\be
\chi_{\mbox{\ssz{imp}}}(T)=\mbox{$\frac{r}{2}$}\chi_{\mbox{\tiny{Curie}}}(T)+c_1'T^{-r}+c_2'T^{-2r}
\ee
\be
C_{\mbox{\ssz{imp}}}(T)\propto T^{1-r}
\ee
\be
S_{\mbox{\ssz{imp}}}(T)=2r\ln2+e'T^{1-r}.
\ee
\reseteqn
It is well known [10,13] that the leading low-$T$ behaviour of $\chi_{\mbox{\ssz{imp}}}$ and $S_{\mbox{\ssz{imp}}}$ for the SC phase---viz $\frac{r}{2}\chi_{\mbox{\tiny{Curie}}}$ and $2r\ln2$ respectively---are given precisely by the $U = 0$ limit result.  But it is striking to note that the leading $T$-dependences of $C_{\mbox{\ssz{imp}}}(T)$ and $\Delta S_{\mbox{\ssz{imp}}}(T)$, namely $T^{1-r}$, are also inherent to the non-interacting limit, as too is the $T^{-r}$ correction to $\chi_{\mbox{\ssz{imp}}}$.  Only the $T^{-2r}$ contribution to the NRG $\chi_{\mbox{\ssz{imp}}}$---which applies only for $r<\frac{1}{2}$ where the SC phase arises for $U>0$---is absent in the non-interacting limit.

\subsection{Local impurity susceptibility}
The `excess' $\chi ^0_{\mbox{\ssz{imp}}}(T)$ discussed above refers to the change in the {\it total} uniform spin susceptibility of the system induced by addition of the impurity.  Here we comment briefly on the {\it local} impurity susceptibility $\chi_{ii}(T)$, which has also been studied via NRG [10,13].  It is defined by
\alpheqn
\be
\chi_{ii}(T)=-g\mu_{\mbox{\ssz{B}}}\frac{\partial\bra\hat{S}_{iz}\ket}{\partial h}\biggm|_{h=0}
\ee
with $\hat{S}_{iz}$ referring to the impurity spin and $h$ a magnetic field acting solely on the impurity; and is given in standard notation by
\be
\chi_{ii}(T)=(g\mu_{\mbox{\ssz{B}}})^2\int_0^\beta \rmd\tau \bra \hat{S}_{iz}(\tau)\hat{S}_{iz} \ket
\ee
\reseteqn
with $\beta = 1/T$.  In the non-interacting limit, $\chi^0_{ii}(T)$ is trivially evaluated and given by
\be
\fl \chi^0_{ii}(T)=\frac{(g\mu_{\mbox{\ssz{B}}})^2}{2}\int_{-\infty}^{\infty}\rmd\omega_1\ \int_{-\infty}^{\infty}\rmd\omega_2 \ \frac {D_0(\omega_1)D_0(\omega_2)}{\omega_1-\omega_2}f(\omega_1)\left[1-f(\omega_2)\right]\left[e^{\beta(\omega_1-\omega_2)}-1\right].
\ee
Separating $D_0(\omega)=q\delta(\omega)+D_0^{\mbox{\ssz{b}}}(\omega)$, and using particle-hole symmetry, a simple calculation gives
\bea
\fl \frac{\chi^0_{ii}(T)}{(g\mu_{\mbox{\ssz{B}}})^2}=\frac{q^2}{8T}&+\frac{q}{2}\int_{-\infty}^{\infty}\rmd\omega\ \frac{D_0^{\mbox{\ssz{b}}}(\omega)}{\omega}\tanh\left(\mbox{$\frac{\omega}{2T}$}\right)\nonumber \\
&+\frac{1}{2}\int_{-\infty}^{\infty}\rmd\omega_1\int_{-\infty}^{\infty}\rmd\omega_2 \ D_0^{\mbox{\ssz{b}}}(\omega_1)D_0^{\mbox{\ssz{b}}}(\omega_2)\frac{\left[f(\omega_2)-f(\omega_1)\right]}{\omega_1-\omega_2}.
\eea

For the gapless LM regime $r>1$, where the poleweight $q(r)\neq 0$ is given by equation (3.9) (with $\delta = 0$), the low-$T$ behaviour of $\chi^0_{ii}(T)$ is thus
\be
\chi^0_{ii}(T)=\mbox{$\frac{q^2}{2}$}\chi_{\mbox{\tiny{Curie}}}(T)\ \ \  :\  r>1
\ee
with leading corrections $\Or\left[\mbox{min}(T^{r-2},1)\right]$ arising from the second term of equation (3.24).  Hence, as expected for a LM state and mirrored also in $\chi^0_{\mbox{\ssz{imp}}}(T)$ equation (3.13a), the impurity spin remains unquenched; although $\chi^0_{ii}(T)/\chi^0_{\mbox{\ssz{imp}}}(T) = q^2 <1$ as $T \rightarrow 0$, reflecting the fact that the `whole' excess level induced by the impurity has only partial weight on the impurity itself.

For the SC phase $r<1$, by contrast, the excess level has no overlap on the impurity, $q=0$.  Only the final term in equation (3.24) survives, and using equation (3.14a) is $\Or(1)$ for $r<\frac{1}{2}$ and $\Or(T^{1-2r})$ for $\frac{1}{2}<r<1$; hence, in particular,
\be
\lim_{T\rightarrow 0}T\chi^0_{ii}(T)=0 \ \ \ : \  0\leq r < 1.
\ee
As for the normal ($r=0$) Anderson model, the impurity spin is thus locally quenched in the entire SC regime $r<1$; in contrast, for obvious reasons, to the behaviour of $\chi^0_{\mbox{\ssz{imp}}}$, equation (3.20a).

The above behaviour---complete (SC) versus incomplete (LM) quenching of the impurity spin---is also found in NRG studies of the soft-gap Kondo model [11,13]; and the total spin quenching symptomatic of the SC phase is one reason why it may be regarded [13] as a natural generalization of conventional Fermi liquid physics.

\ \ \newline
We have seen that even the non-interacting limit contains both LM and SC states, whose characteristics are reflected in the behaviour of the impurity spectrum $D_0(\omega)$---by whether (LM) or not (SC) there is a $\delta$-function contribution at the Fermi level---and hence in turn in $\Delta\rho_0(\omega)$ and resultant thermodynamic functions.  One final point should however be noted: the $U = 0$ `phase diagram' consists of LM states for $r>1$ and SC states for $r<1$.  This is in contrast to what is found by NRG for $U > 0$ [12,13], where SC states arise only for $r<\frac{1}{2}$ and where only LM states occur for $r > \frac{1}{2}$.  We consider the implications of this further in section 5.
\seceq
\section{$U>0$:  conditions for SC phase}
For non-vanishing interaction strengths, $U$, we now seek conditions under which, upon increasing $U$ from zero, a SC state will persist; (only the gapless problem, and for $r<1$, need be considered, since for $r>1$ the $U=0$ ground state is a local moment one).  Employing a phase shift analysis that parallels Nozi\`{e}res' Fermi liquid description of the `normal' Kondo effect [15], Chen and Jayaprakash [10] have argued that the SC state is characterized generally by
\be
\Delta\rho^{\mbox{\ssz{p}}}(\omega)=r\delta(\omega) \ \ \ \ \ \:\ \ \ :\mbox{SC}
\ee
---precisely as for the non-interacting limit.

The question now is:  under what conditions upon the self-energy $\Sigma(\omega)$ will this arise? $\Delta\rho^{\mbox{\ssz{p}}}(\omega)$ is given by equation (2.10b) with $D^{\mbox{\ssz{p}}}(\omega)$ from equation (2.8b).  Since $\Delta_{\mbox{\ssz{I}}}$ and  $\Sigma_{\mbox{\ssz{I}}}$ are non-negative and, for $|\omega|<D$, $\Delta_{\mbox{\ssz{I}}}(\omega)=\Delta_0|\omega|^r$ vanishes only for $\omega = 0$, the only $\delta$-funtion contribution that could arise in $\Delta\rho^{\mbox{\ssz{p}}}(\omega)$ for $|\omega|<D$ is of course for $\omega = 0$.  Hence only the low-$\omega$ behaviour of $\Delta$ and $\Sigma$ is relevant, and  $\Delta\rho^{\mbox{\ssz{p}}}(\omega)$ is given by (cf equation(3.19a))
\alpheqn
\be
\Delta\rho^{\mbox{\ssz{p}}}(\omega)=D^{\mbox{\ssz{p}}}(\omega)\left[1+\Delta_0r\tan\left(\mbox{$\frac{\pi }{2}$}r\right)|\omega|^{r-1}\right]
\ee
with
\be
D^{\mbox{\ssz{p}}}(\omega)\equiv\frac{\eta\pi^{-1}}{\left[\Delta_{\mbox{\ssz{R}}}(\omega)+\Sigma_{\mbox{\ssz{R}}}(\omega)\right]^2+\left[\eta+\Delta_{\mbox{\ssz{I}}}(\omega)+\Sigma_{\mbox{\ssz{I}}}(\omega)\right]^2}
\ee
\reseteqn
(where the `bare' $\omega$ contribution to equation (2.8b) can again be neglected since, for $r<1$, it is subdominant to $\Delta_{\mbox{\ssz{R}}}(\omega)\sim|\omega|^r$).  Notice also that if the low-$\omega$ behaviour of $\Sigma_{\mbox{\ssz{I}}}(\omega)$ is of form
\alpheqn
\be
\Sigma_{\mbox{\ssz{I}}}(\omega)=\alpha|\omega|^{\lambda} \ \ \ :\ \ \ \omega\rightarrow 0
\ee
with $-1<\lambda<1$, where $\alpha\equiv\alpha(U)$ is a (necessarily positive) constant, then the corresponding low-$\omega$ behaviour of $\Sigma_{\mbox{\ssz{R}}}(\omega)$ follows directly from the Hilbert transform equation (2.6) as
\be
\Sigma_{\mbox{\ssz{R}}}(\omega)=-\mbox{sgn}(\omega)\tan\left(\mbox{$\frac{\pi}{2}\lambda$}\right)\Sigma_{\mbox{\ssz{I}}}(\omega) \ \ \ \ :\ \ \ \ \omega \rightarrow 0.
\ee
\reseteqn

If $\Sigma_{\mbox{\ssz{I}}}(\omega)$ and hence  $\Sigma_{\mbox{\ssz{R}}}(\omega)$ decay to zero as $\omega \rightarrow 0$ more rapidly than  $\Delta_{\mbox{\ssz{I}}}(\omega)$---i.e. if  $\Sigma_{\mbox{\ssz{I}}}(\omega)$ is of form equation (4.3a) with $r<\lambda$---then, trivially, the low-$\omega$ behaviour of equation (4.2b) for $D^{\mbox{\ssz{p}}}(\omega)$ is precisely that of the $U = 0$ limit, i.e. $D^{\mbox{\ssz{p}}}(\omega) = L(\Delta_{\mbox{\ssz{I}}};r)$ (see equation (3.16)).  Hence, as in equations (3.17) ff, $\Delta\rho^{\mbox{\ssz{p}}}(\omega) = r\delta(\omega)$ arises.  If by contrast $\Sigma_{\mbox{\ssz{I/R}}}$ dominate the low-$\omega$ behaviour of equation (4.2b)---i.e. if $\Sigma_{\mbox{\ssz{I}}}(\omega)$ is of form equation (4.3a) with $\lambda < r$---then from equations (4.3) and (3.17), $D^{\mbox{\ssz{p}}}(\omega) = L(\Sigma_{\mbox{\ssz{I}}};\lambda)$; from equation (4.2a) it then follows that $\Delta\rho^{\mbox{\ssz{p}}}(\omega)\sim |\omega|^{r-\lambda}\delta(\omega)\equiv 0$ since $\lambda < r$, i.e. there is no pole contribution to $\Delta\rho(\omega)$.  Finally, if the low-$\omega$ behaviour of $\Sigma_{\mbox{\ssz{I/R}}}$is the same as that of $\Delta_{\mbox{\ssz{I/R}}}$, viz $\lambda = r$ in equations (4.3), then a directly analogous calculation gives $\Delta\rho^{\mbox{\ssz{p}}}(\omega)=r/[1+\alpha(U)/\Delta_0]\ \delta(\omega)$; i.e. a $\delta$-function contribution but with a $U$-dependent weight that is less than $r$.

The question posed above is thus answered:  for $\Delta\rho^{\mbox{\ssz{p}}}(\omega) = r\delta(\omega)$ to arise, and thus a SC state to be realized for $U>0$, $\Sigma_{\mbox{\ssz{I}}}(\omega)$ and hence $\Sigma_{\mbox{\ssz{R}}}(\omega)$ must decay to zero as $\omega \rightarrow 0$ {\it more} rapidly that $|\omega|^r$, i.e.
\be
\Sigma_{\mbox{\ssz{I}}}(\omega)\stackrel{\omega \rightarrow 0}{\sim}\alpha|\omega|^{\lambda}\ \ \ \ :\ \ \ \ r<\lambda
\ee

This has important implications for the low-$\omega$ behaviour of the impurity single-particle spectrum $D(\omega)$ ($\equiv D^{\mbox{\ssz{b}}}(\omega)$), since from equation (2.8a) it follows that the low-$\omega$ asymptotic behaviour of $D(\omega)$ is precisely that of $D_0(\omega)$, whence in particular
\be
D(\omega) \stackrel{\omega \rightarrow 0}{\sim}|\omega|^{-r}
\ee
as indeed found for the SC phase in NRG studies of the impurity spectrum [12].  More importantly, using
\be
\lim_{\omega \rightarrow 0} |\omega|^rD(\omega) = \lim_{\omega \rightarrow 0} |\omega|^rD_0(\omega)
\ee
together with equation (3.14), and defining $A(\omega)=|\omega|^rD(\omega)$, it follows that 
\be
\pi\Delta_0 \left[1+\tan^2\left(\mbox{$\frac{\pi}{2}r$}\right) \right]A(\omega = 0)=1.
\ee
We believe this to be significant.  For the `normal' Anderson model, $r=0$, it recovers the well known result (see e.g.[1]), normally viewed as a consequence of the Friedel sum rule, that the impurity single particle spectrum is pinned at the Fermi level $\omega = 0$, i.e. that $D(\omega = 0) = 1/\pi \Delta_0$ (in this case for all $U$). Equation (4.7) represents a generalization of this pinning condition to arbitrary $r$ where a SC state obtains, whose continuity in $r$ reflects the fact that interactions have no influence in renormalizing the asymptotic behaviour of $D(\omega)$ as $\omega \rightarrow 0$; and which is entirely consistent with the conlusions of Gonzalez-Buxton and Ingersent [13] from NRG studies that the SC state embodies a natural generalization of standard Fermi liquid physics.  The extent to which equation (4.7) is captured in practice should also provide a good test of the accuracy of NRG calculations at low frequencies, and will be discussed elsewhere; moreover the generalized pinning condition will prove central to our local moment approach to the problem, as will be discussed in a subsequent paper [14].

One important question is not of course answered by the above considerations: for what range of $r$ will the condition equation (4.4) for a SC state actually arise?  NRG calculations give $r < \frac{1}{2}$ for the SC state [10-13].  We examine this question in the following section within the framework of second order PT in $U$, together with a corresponding analysis of the evolution with $U$ of the single-particle spectra appropriate to the LM state.

\seceq
\section{Perturbation theory in $U$}
Low order perturbation theory in $U$ about the non-interacting limit is probably the simplest and certainly the most conventional approach to the problem.  For the `normal' Anderson model, $r=0$, its predictions are of course well known (see e.g.[1]):  while restricted by contruction to weak coupling interactions $U$, and thus incapable of capturing strong coupling `Kondo' asymptotics, it generates order by order characteristic Fermi liquid behaviour, in particular that $\Sigma_{\mbox{\ssz{I}}}\sim \Or(\omega^2)$ and that the impurity spectrum is pinned at the Fermi level, $D(0) = 1/\pi\Delta_0$; and the single-particle spectrum evolves continuously upon increasing $U$ from the non-interacting limit, in accordance with the fact that the normal Anderson model is a Fermi liquid for all $U/\Delta_0 \geq 0$.  For $r>0$ by contrast the implications of a low order perturbative treatment---and even whether such is in general applicable---are far from obvious, and are considered here at the simplest second order level.
\restylefloat{figure}
\begin{figure}[H] 
\centering 
\epsfig{file =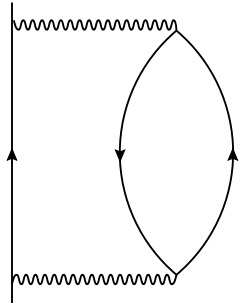,width=3cm} 
\caption{Second order self-energy diagram, with bare ($U=0$) impurity propagators denoted by solid lines and the on-site impurity $U$ by wavy lines.}
\end{figure}

The familiar second order self-energy diagram is shown in figure 2 and may be written as
\be
\Sigma(\omega)=U^2\int_{-\infty}^{\infty}\frac{\rmd\omega_1}{2\pi i}\ ^0\!\Pi(\omega_1)G_0(\omega_1+\omega)
\ee
where $G_0(\omega)$ is the non-interacting impurity Green function with spectral representation
\be
G_0(\omega)=\int_{-\infty}^{\infty}\rmd\omega_1\frac{D_0(\omega_1)}{\omega-\omega_1+i\eta \mbox{sgn}(\omega)}.
\ee
\alpheqn
The `polarization bubble' $\ ^0\!\Pi(\omega)$ is given by
\bea
\ ^0\!\Pi(\omega)&=i\int_{-\infty}^{\infty}\frac{\rmd\omega_1}{2\pi}G_0(\omega_1)G_0(\omega_1-\omega)\\
		&=\ ^0\!\Pi(-\omega)
\eea
\reseteqn
(where the latter follows from a trivial change of variables in equation (5.3a)); its spectral representation is
\be
\ ^0\!\Pi(\omega)=\int_{-\infty}^{\infty}\frac{\rmd\omega_1}{\pi}\frac{\mbox{Im}\ ^0\!\Pi(\omega_1)\mbox{sgn}(\omega_1)}{\omega_1-\omega-i\eta\mbox{sgn}(\omega)}.
\ee

From equation (5.1), using (5.2) and (5.4) together with $D_0(\omega)=D_0(-\omega)$ (particle-hole symmetry), a straightforward calculation gives
\be
\Sigma_{\mbox{\ssz{I}}}(\omega)=U^2\int_0^{|\omega|}d\omega_1\mbox{Im}\ ^0\!\Pi(\omega_1)D_0(\omega_1-|\omega|)
\ee
which is thus readily calculable from a knowledge of $D_0(\omega)$ (section 3) and $\ ^0\!\Pi(\omega)$.  Similarly, using equation (5.2) again, a directly analogous calculation yields
\be
\mbox{$\frac{1}{\pi}$}\mbox{Im}\ ^0\!\Pi(\omega)=\int_0^{|\omega|}\rmd\omega_1 D_0(\omega_1)D_0(\omega_1-|\omega|)\ \ \geq 0.
\ee
Note that, using equation (5.6), equation (5.5) implies $\Sigma_{\mbox{\ssz{I}}}(\omega)\geq 0$ as required by analyticity.  Equations (5.5) and (5.6) are the basic equations to be analyzed, as now considered for the gapless problem ($\delta = 0$).

Separating $D_0(\omega)$ into pole and band contributions, and using $D_0^{\mbox{\ssz{p}}}(\omega)=q\delta(\omega)$, equation (5.6) reduces for $|\omega| < D$ to
\be
\mbox{$\frac{1}{\pi}$}\mbox{Im}\ ^0\!\Pi(\omega)=\mbox{$\frac{1}{2}$}q^2\delta(\omega)+qD_0^{\mbox{\ssz{b}}}(\omega)+\int_0^{|\omega|}\rmd\omega_1D_0^{\mbox{\ssz{b}}}(\omega_1)D_0^{\mbox{\ssz{b}}}(\omega_1-|\omega|).
\ee
This encompasses both $r>1$ where $q \neq 0$ is given from equation (3.9) with $\delta= 0$; and $r<1$ where $q = 0$ (section 3.3) and only the `band-band' contribution  to $\mbox{Im}\ ^0\!\Pi(\omega)$ survives.  [We add in passing that pole-contributions to $D_0^{\mbox{\ssz{p}}}(\omega)$ from outside the band, $|\omega|>D$, generate spectral contributions to $\mbox{Im}\ ^0\!\Pi(\omega)$ for $D<|\omega|<2D$; as we shall be concerned only with the low-$\omega$ behaviour of  $\mbox{Im}\ ^0\!\Pi(\omega)$ and hence $\Sigma_{\mbox{\ssz{I}}}(\omega)$, we refrain from showing these, although they are fully included in numerical calculations where they are necessary to ensure the correct normalization of $D(\omega)$.]

The low-$\omega$ asymptotics of  $\mbox{Im}\ ^0\!\Pi(\omega)$ follow from equation (5.7) using the results of section 3 for $D_0(\omega)$.  For $r>1$ we obtain
\be
\mbox{$\frac{1}{\pi}$}\mbox{Im}\ ^0\!\Pi(\omega)\stackrel{\omega \rightarrow 0}{\sim}\mbox{$\frac{1}{2}$}q^2\delta(\omega)+qD_0^{\mbox{\ssz{b}}}(\omega)+\Or(|\omega|^{2r-3})\ \ \ :\ \ \ r>1
\ee
where the $\Or(|\omega|^{2r-3})$ corrections arise from the band-band piece of equation (5.7) and diverge less rapidly as $\omega \rightarrow 0$ than the leading $|\omega|^{r-2}$ behaviour of $D_0^{\mbox{\ssz{b}}}(\omega)$ (equation (3.11)).  For $r<1$ by contrast, only the band-band contribution to equation (5.7) survives, and a simple calculation using equation (3.14) gives the leading low-$\omega$ behaviour
\alpheqn
\be
\mbox{$\frac{1}{\pi}$}\mbox{Im}\ ^0\!\Pi(\omega)\stackrel{\omega \rightarrow 0}{\sim}C(r)|\omega|^{1-2r}\ \ \ \ :\ \ \ \ r < 1
\ee
with $C>0$ given by
\bea
C(r)=&\frac{\cos^4\left(\frac{\pi}{2}r\right)}{\pi\Delta_0}\int_0^1\rmd y\ y^{-r}(1-y)^{-r}\\
=&\frac{\cos^4\left(\frac{\pi}{2}r\right)}{\Delta_0}\frac{2^{2r-1}}{\sqrt{\pi}}\frac{\Gamma(1-r)}{\Gamma(\frac{3}{2}-r)}.
\eea
\reseteqn
These results may now be used in equation (5.5) to determine the crucial low-$\omega$ asymptotics of $\Sigma_{\mbox{\ssz{I}}}(\omega)$, as now considered separately for $r>1$ and $r<1$.

\subsection{$r>1$: LM state}
From equations (5.5) and (5.8) the low-$\omega$ behaviour of $\Sigma_{\mbox{\ssz{I}}}(\omega)$ is given by
\alpheqn
\bea
\Sigma_{\mbox{\ssz{I}}}(\omega)&=\frac{\pi U^2q^2}{4}\left[q\delta(\omega)+3D_0^{\mbox{\ssz{b}}}(\omega)+\Or(|\omega|^{2r-3})\right]\\
		&=\frac{\pi U^2 q^3}{4}\delta(\omega)+\Sigma_{\mbox{\ssz{I}}}^{\mbox{\ssz{b}}}(\omega)
\eea
where, using equation (3.11), the asymptotic behaviour of the `band' contribution $\Sigma_{\mbox{\ssz{I}}}^{\mbox{\ssz{b}}}(\omega)$ is
\be
\Sigma_{\mbox{\ssz{I}}}^{\mbox{\ssz{b}}}(\omega)\stackrel{\omega \rightarrow 0}{\sim}\frac{3}{4}U^2q^4\Delta_0|\omega|^{r-2}.
\ee
\reseteqn
The low-$\omega$ behaviour of $\Sigma_{\mbox{\ssz{I}}}(\omega)$ is thus integrably singular, and manifestly not that of a Fermi liquid or any natural generalization thereof: not surprisingly, since the underlying $U=0$ ground state for $r>1$ is a local moment one.  The corresponding real part, $\Sigma_{\mbox{\ssz{R}}}(\omega)$, follows from the Hilbert transform equation (2.6) using equation (5.10a), and has the leading low-$\omega$ behaviour
\be
\Sigma_{\mbox{\ssz{R}}}(\omega)\stackrel{\omega \rightarrow 0}{\sim}\frac{U^2q^3}{4}P\left(\frac{1}{\omega}\right)
\ee
(with corrections $\Or(|\omega|^{r-2},|\omega|)$ arising from the transform of  $\Sigma_{\mbox{\ssz{I}}}^{\mbox{\ssz{b}}}(\omega)$).

From equation (2.8) the low-$\omega$ behaviour of the impurity spectrum $D(\omega)$ thus follows directly as
\be
D(\omega)\stackrel{\omega \rightarrow 0}{\sim}\frac{\Sigma_{\mbox{\ssz{I}}}^{\mbox{\ssz{b}}}(\omega)}{\pi\left[\Sigma_{\mbox{\ssz{R}}}(\omega)\right]^2} = \frac{12}{\pi}\frac{\Delta_0}{(Uq)^2}|\omega|^r.
\ee    
This behaviour---$D(\omega)\sim |\omega|^r$---is precisely the spectral hallmark of the LM regime found in NRG calculations for $U>0$ [12].  It is of course in marked contrast to what obtains in the non-interacting limit, viz equation (3.11) $D_0^{\mbox{\ssz{b}}}(\omega)\sim |\omega|^{r-2}$ (together with a pole contribution for $U = 0$, which is eliminated entirely for $U>0$).  Although the lowest frequency spectral asymptotics thus change character abruptly upon increasing $U$ from zero, it is however straightforward to show (using the asymptotic forms of $\Sigma_{\mbox{\ssz{R}}}/\Sigma_{\mbox{\ssz{I}}}^{\mbox{\ssz{b}}}$ above) that for sufficiently small $U$ there exists a crossover scale $\omega_0 = \frac{1}{2}Uq^2$ such that for ($D \gg$)$|\omega|\gg\omega_0$ the behaviour of $D(\omega)$ is that of the non-interacting limit, viz $D(\omega)\sim|\omega|^{r-2}$; while for $|\omega|\ll\omega_0$, $D(\omega)\sim |\omega|^r$ as in equation (5.12).

Representative single-particle spectra for the LM state, obtained at the second-order level, are illustrated in figures 3 and 4 for $r = 1.5$.  We consider first a `weak hybridization' example which, for $r>1$, entails $\tilde{D}=\Delta_0^{\frac{1}{r-1}}D\ll1$ (see equation (2.12)).  For $\tilde{D} = 1.5 \times 10^{-3}$, figure 3a shows the dimensionless $D'(\tilde{\omega})=\Delta_0^{\frac{1}{1-r}}D(\omega)$ vs. $\tilde{\omega}=\omega/\Delta_0^{\frac{1}{1-r}}$ for three reduced interaction strengths $\tilde{U} = U/\Delta_0^{\frac{1}{1-r}} = 2.5 \times 10^{-4}$, $5\times 10^{-4}$ and $7.5\times 10^{-4}$.  The dominant visible feature of the spectra are the Hubbard satellites, which for all $\tilde{U}$'s shown are centred to high accuracy on $\tilde{\omega_0} = \frac{1}{2}\tilde{U}q^2$ (with the poleweight $q \approx 0.95$ from equation (3.9) with $\delta=0$).  This is expected physically:  the weak hybridization regime is `close' to the atomic limit, $\Delta_0 = 0 = V_{i \bi{k}}$ (where $q=1$), which by a well known accident for the particle-hole symmetric case (see e.g. [1]) is captured exactly by second order PT, and where equation (5.11) with $q=1$ is exact for all $\omega$.  The sharp Hubbard satellites of figure 3a thus correspond simply to weak resonant broadening, and only slight shifting, of the atomic limit poles occuring at $\omega = \pm \frac{1}{2}U$.  The low-$\omega$ behaviour of the spectrum, equation (5.12), is not directly visible in figure 3a, but is clear from the inset to figure 3a and in figure 3b.  In the latter, for the same parameters as figure 3a, we show $\log \left[(\pi[\tilde{U}q]^2/12)D'(\tilde{\omega})\right]$ vs. $\log \tilde{\omega}$.  That $D'(\tilde{\omega})\sim |\tilde{\omega}|^r$ as $\tilde{\omega}\rightarrow 0$ is evident; as too is the accuracy of equation (5.12) in its entirety.  The above mentioned crossover to $|\tilde{\omega}|^{r-2}$ behaviour for $|\tilde{\omega}|\gg \tilde{\omega}_0$ is also evident in figure 3b for the lowest $\tilde{U}$ example.

\restylefloat{figure}
\afterpage{\begin{figure}[H] 
\centering 
\epsfig{file =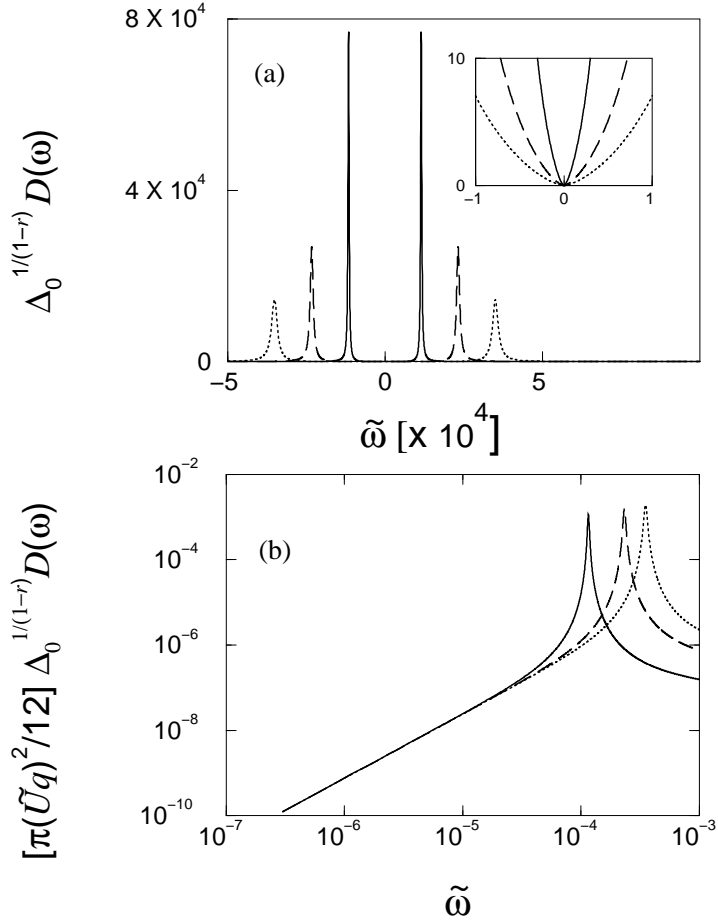,width=10cm} 
\caption{LM regime, $r=1.5$. (a)$D'(\tilde{\omega})=\Delta_0^{\frac{1}{1-r}}D(\omega)$ versus $\tilde{\omega}=\omega/\Delta_0^{\frac{1}{1-r}}$ for reduced interaction strengths $\tilde{U} = 2.5\times 10^{-4}$ (\full), $5\times 10^{-4}$ (\dashed) and $7.5\times 10^{-4}$ (\dotted); all for reduced bandwidth $\tilde{D} = 1.5\times 10^{-3}$.  Inset: low-$\tilde{\omega}$ behaviour of $D'(\tilde{\omega})$.  (b) $\left[\pi(\tilde{U}q)^2/12\right]D'(\tilde{\omega})$ versus $\tilde{\omega}$ on a log-log scale for same parameters as (a).}
\end{figure}}

The spectral features naturally evolve smoothly with increasing hybridization strength.  For $r=1.5$ again, figure 4 shows $D'(\tilde{\omega})$ vs. $\tilde{\omega}$ for a `strong hybridization' example, $\tilde{D} = 10$ (where $q \approx0.2 \ll 1)$, for $\tilde{U} = 4$, 8 and 20.  The low-$\tilde{\omega}$ behaviour, $D'(\tilde{\omega})\sim |\tilde{\omega}|^r$, is clearly seen in all cases.  The Hubbard satellites are again centred on $\tilde{\omega}_0=\frac{1}{2}\tilde{U}q^2$ for sufficiently low $\tilde{U}$ (e.g. for $\tilde{U}=4$); and in all cases, since $q \ll 1$, lie well below the scale of $\frac{1}{2}\tilde{U}$ characteristic of the atomic limit.  It is also seen that the satellites become increasingly diffuse with increasing interaction strength, although whether this is a genuine feature is not clear since second order PT is by construction confined to weak coupling.

\restylefloat{figure}
\afterpage{\begin{figure}[H] 
\centering 
\epsfig{file =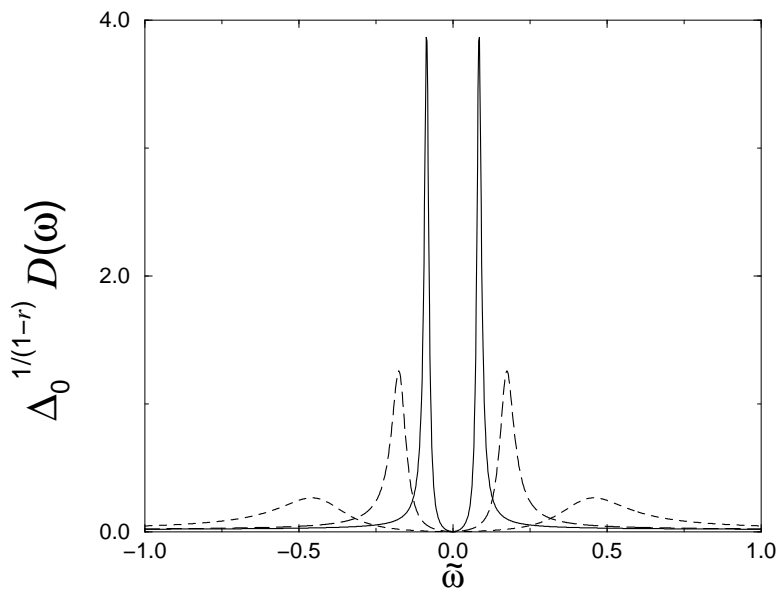,width=10cm} 
\caption{LM regime, $r = 1.5$ and reduced bandwidth $\tilde{D}=10$.  $D'(\tilde{\omega})$ versus $\tilde{\omega}$ for $\tilde{U}$ = 4 (\full), 8 (\broken) and 20 (\dashed).}
\end{figure}}

\subsection{$0\leq r<\frac{1}{2}$: SC state}
For $r<1$, $\mbox{Im}\ ^0\!\Pi(\omega)$ is given by equation (5.9) and the $\omega \rightarrow 0$ behaviour of $D_0(\omega)(\equiv D_0^{\mbox{\ssz{b}}}(\omega))$ by equation (3.14).  Hence from equation (5.5) the low frequency behaviour of $\Sigma_{\mbox{\ssz{I}}}(\omega)$ is
\alpheqn
\be
\Sigma_{\mbox{\ssz{I}}}(\omega)\stackrel{\omega \rightarrow 0}{\sim}\alpha(U)|\omega|^{2-3r}
\ee
with
\bea
\alpha(U)&=U^2\frac{\cos^2\left(\frac{\pi}{2}r\right)}{\Delta_0}C(r)\int_0^1\rmd y\ y^{1-2r}(1-y)^{-r} \nonumber\\
&=U^2\frac{\cos^2\left(\frac{\pi}{2}r\right)}{\Delta_0}C(r)\frac{\Gamma(2[1-r])\Gamma(1-r)}{\Gamma(3[1-r])}
\eea
\reseteqn
and $C(r)$ given by equation (5.9).  From the Hilbert transform equation (2.6), the corresponding behaviour of $\Sigma_{\mbox{\ssz{R}}}(\omega)$ follows as
\begin{displaymath}
 \Sigma_{\mbox{\ssz{R}}}(\omega)\stackrel{\omega \rightarrow 0}{\sim}
\left\{\begin{array}{ccc}
-\mbox{sgn}(\omega)\gamma|\omega|\ \ \ \ \ \ \ \ \ \ \ \ \ \ \ \ \ \ 
 \ \ \ \ \ \ \ \ \ \ \   &:& 0 \leq r < \frac{1}{3} \mbox{\ \ (5.14a)}\\
-\mbox{sgn}(\omega)\tan\left[\frac{\pi}{2}(2-3r)\right]\alpha(U)|\omega|^{2-3r}&:&r>\frac{1}{3} \mbox{\ \ \ \ \ \ \ (5.14b)}\\ 
\end{array}
\right.\\
\end{displaymath}
\setcounter{equation}{14}
(with logarithmic behaviour $\sim |\omega|\ln|\omega|$ for $r=\frac{1}{3}$), where $\pi \gamma = \int_{-\infty}^{\infty}\rmd\omega\ \Sigma_{\mbox{\ssz{I}}}(\omega)/\omega^2 >0$.

The important point here is that for $r< \frac{1}{2}$, where $r<2-3r$, $\Sigma_{\mbox{\ssz{I}}}(\omega)$ and $\Sigma_{\mbox{\ssz{R}}}(\omega)$ decay to zero as $\omega \rightarrow 0$ more rapidly than $\Delta_{\mbox{\ssz{I/R}}}\sim |\omega|^r$.  The conditions established in section 4 for the SC state are thus satisfied at the level of second order perturbation theory, and the SC state is hence perturbatively continuable from the non-interacting limit.  Interactions do not renormalize the asymptotic behaviour of $D(\omega)$ as $\omega \rightarrow 0$, and the generalized pinning condition equation (4.7) is thus satisfied.  Note moreover that, as for the normal Anderson model $r=0$, $\Sigma_{\mbox{\ssz{I}}}(\omega)$ vanishes at the Fermi level $\omega = 0$.

Representative single-particle spectra for the SC phase are shown in figure 5 which, for $r = \frac{1}{4}$ and the wide-band limit ($D=\infty$), shows $D'(\tilde{\omega})$ vs. $\tilde{\omega}$ for $\tilde{U} = $ 1, 5 and 7.5.  As found in NRG studies [12], the gross features of the spectrum are dominated by the low-$\omega$ behaviour where $D(\omega)\sim |\omega|^{-r}$.  And with increasing interaction strength $\tilde{U}$, the low frequency resonant continuum narrows and emergent Hubbard satellites become evident as weak shoulders in the spectrum.

\restylefloat{figure}
\begin{figure}[H] 
\centering 
\epsfig{file =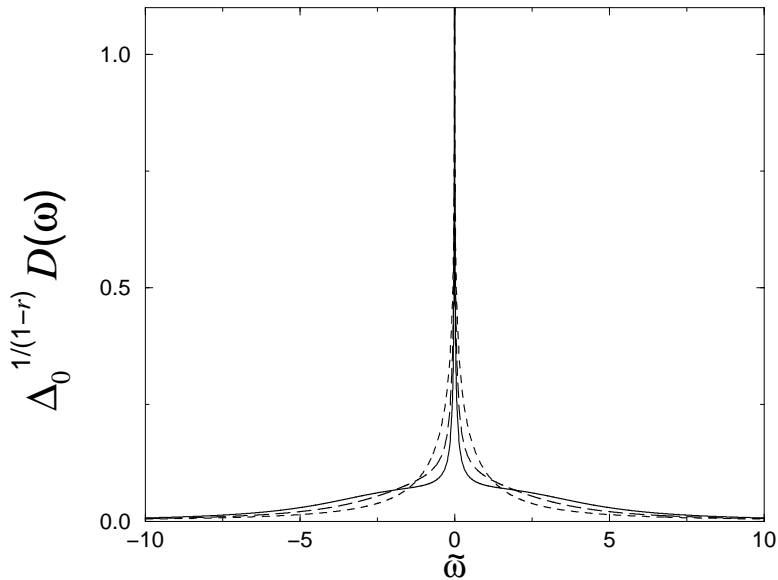,width=10cm} 
\caption{SC phase, $r=\frac{1}{4}$. $D'(\omega)=\Delta_0^{\frac{1}{1-r}}D(\omega)$ versus $\tilde{\omega}$ for $\tilde{U}$ = 1 (\dashed), 5 (\broken) and 7.5 (\full).}
\end{figure}

A far more revealing expos\'{e} is seen in figure 6 where, for the same parameters as figure 5, $\pi \Delta_0\left[1+\tan^2\left(\frac{\pi}{2}r\right)\right]A(\omega)$ vs. $\tilde{\omega}$ is shown, with
\be
A(\omega)=|\omega|^rD(\omega).
\ee
Note first that the generalized pinning condition equation (4.7), which applies for any $r$ where a SC state obtains, is manifestly satisfied.  The low-$\omega$ behaviour of $A(\omega)$ is cusp-like, and from the asymptotics given above it is readily shown that $A(\omega)\sim A(0)-\left[a_1|\omega|^{1-r}+a_2|\omega|^{2(1-2r)}\right]$ where the coefficient $a_1 \propto \tan (\pi r/2)$.  For $0<r<\frac{1}{3}$ the cusp is thus of form $\Delta A(\omega)\sim |\omega|^{1-r}$, while for $\frac{1}{3}<r<\frac{1}{2}$ it behaves as $\Delta A (\omega) \sim |\omega|^{2(1-2r)}$; and for $r=0$, the parabolic behaviour $\Delta A (\omega) \sim \omega^2$ characteristic of `normal' Fermi liquid behaviour is recovered.  In fact, the behaviour of $A(\omega)$ shown in figure 5 for $r=0.25$ is strongly reminiscent of spectra characteristic of the normal Anderson model, $r=0$; the latter being shown in figure 6 for the same $\tilde{U}$ values.  The parallels are obvious, in each case $A(\omega = 0)$ being $\tilde{U}$-independent (pinned) and with Hubbard satellite peaks progressively evolving with increasing interaction strength.  Most significantly, we see in either case the emergence with increasing $\tilde{U}$ of a low-energy scale, reflected in the half-width, $\omega_K$, of $A(\omega)$ which narrows progressively as $\tilde{U}$ is increased.  For $r=0$ this is just the emergence of the ordinary Kondo scale, while for $0<r<\frac{1}{2}$ it is the generalization thereof known in a thermodynamic context for the soft-gap Kondo model itself [6,10].

\restylefloat{figure}
{\begin{figure}[H] 
\centering 
\epsfig{file =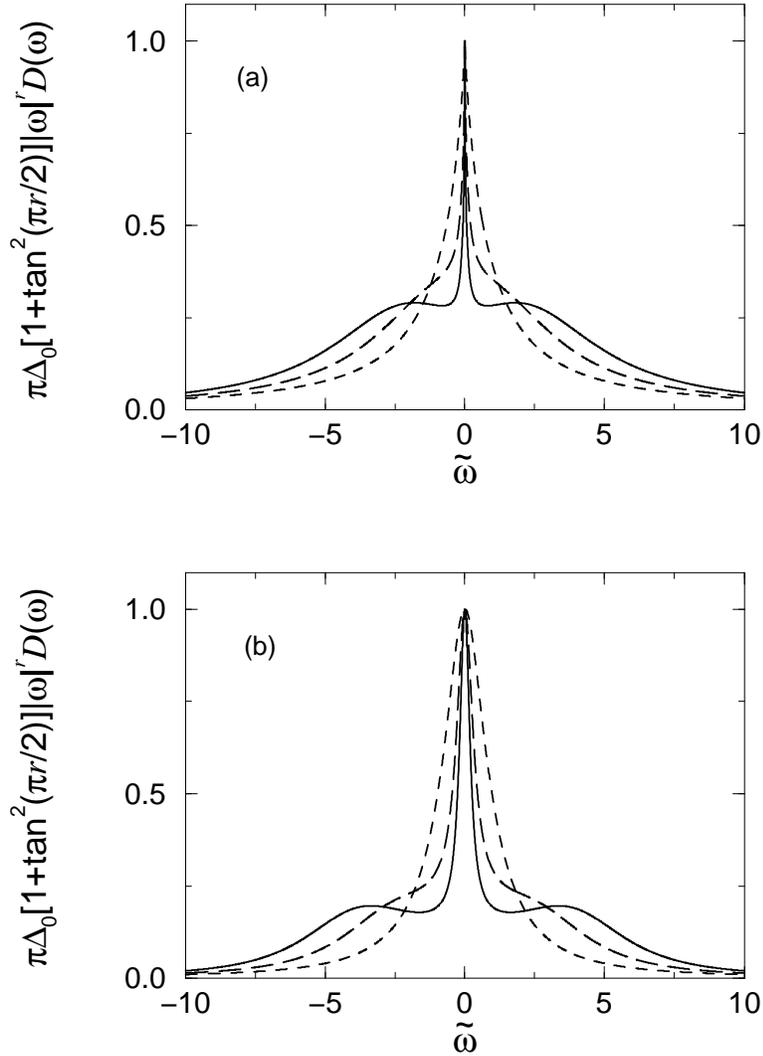,width=10cm} 
\caption{$\pi\Delta_0[1+\tan^2(\frac{\pi}{2}r)]|\omega|^rD(\omega)$ versus $\tilde{\omega}$ for (a) SC phase, $ r = \frac{1}{4}$ and (b) `normal' $r=0$ Anderson model.  For $\tilde{U}$ = 1 (\dashed), 5 (\broken) and 7.5 (\full) in either case.}
\end{figure}

We shall not however pursue here the evolution of the Kondo scale with increasing $\tilde{U}$ since the above analysis, while showing that the SC state is perturbatively continuable from $\tilde{U}=0$ and enabling a description of the spectrum at low $\tilde{U}$, also points clearly to the limitations of a low-order perturbative treatment.  For the normal Anderson model, we know that low-order perturbation theory is intrinsically incapable of describing strong coupling `Kondo' asymptotics---$D(\omega)$ narrowing algebraically with increasing $\tilde{U}$ and failing thereby to yield the correct exponentially small Kondo scale $\omega_K \sim \exp(-\pi U/8\Delta_0)$ (see e.g. [1]).  And the situation is even more acute for the SC phase with $r>0$: here, from NRG studies of the gapless Anderson model [12,13], it is known that for $0<r<\frac{1}{2}$ there is a critical line $\tilde{U}_C(r)$ below which a SC state obtains, and above which the ground state is by contast a LM one.  As $\tilde{U} \rightarrow \tilde{U}_C(r)-$, the low energy Kondo scale $\omega_K$ characteristic of the SC state must vanish, and this clearly cannot be captured by straight  perturbation theory.  To capture such behaviour, and hence in particular to describe the SC/LM phase boundary, an intrinsically non-perturbative approach is needed, and will be given in a subsequent paper [14].
\subsection{$\frac{1}{2}<r<1$}
Finally, we turn to the regime $\frac{1}{2}<r<1$.  Here the low-$\omega$ asymptotics of the second order self-energy are again precisely those given in section 5.2, viz equations (5.13) and (5.14b).  But in contrast to $0\leq r <\frac{1}{2}$, $\Sigma_{\mbox{\ssz{I}}}(\omega)$ and $\Sigma_{\mbox{\ssz{R}}}(\omega)$ do not decay to zero as $\omega \rightarrow 0$ more rapidly than $\Delta_{\mbox{\ssz{I/R}}} \sim |\omega|^r$, whence the conditions established in section 4 for the SC state are {\it not} satisfied for $\frac{1}{2}<r<1$:  there is no pole contribution to $\Delta \rho(\omega)$, and the pinning condition equation (4.7) is not satisfied since the low-$\omega$ behaviour of the impurity spectrum is apparently dominated by $\Sigma_{\mbox{\ssz{I/R}}}$, viz $D(\omega) \sim [\Sigma_{\mbox{\ssz{I}}}(\omega)]^{-1}\sim |\omega|^{3r-2}$---diverging as $\omega \rightarrow 0$ for $\frac{1}{2}<r<\frac{2}{3}$, but less rapidly than the $|\omega|^{-r}$ behaviour characteristic of the SC state; and vanishing for $\frac{2}{3}<r<1$, but less rapidly than the $|\omega|^r$ behaviour symptomatic of the LM state.  Superficially, therefore, one appears confronted by a state that is neither SC or LM in character.

There is, however, a plausible explanation: that straight perturbation theory  in $U$ about the non-interacting limit is intrinsically inapplicable for $\frac{1}{2}<r<1$.  We believe this to be the case, for we have shown in section 3 that for $U = 0$ the ground state is SC for all $r<1$, and LM for $r>1$.  By contrast, finite-$U$ NRG studies [12,13] yield a LM ground state for $r>\frac{1}{2}$, with no indication of a transition between LM and SC states at any finite interaction strength.  For $\frac{1}{2}<r<1$, therefore, the transition between SC and LM states occurs `at' $U=0$ itself, and since the non-interacting and $U>0$ ground states are of different symmetry, one anticipates a breakdown of naive perturbation theory about the non-interacting limit.  For $r<\frac{1}{2}$  or $r>1$ by contrast, the $U=0$ ground states are the same as those for $U>0$ (at least for sufficiently small $U$ in the case of $0<r<\frac{1}{2}$).  Hence, as found in sections 5.1 and 5.2, one expects low order perturbation theory in $U$ to be applicable over some finite-$U$ interval; although we naturally anticipate such to become `dangerous' as $r \rightarrow \frac{1}{2}-$ or $r \rightarrow  1+$.

\seceq
\section{Conclusion}
A summary of the present work is aptly illustrated by figure 7 which, for the symmetric soft-gap Anderson model, shows schematically the phase boundaries between strong coupling (SC) and local moment (LM) phases in the $\tilde{U}=U/\Delta_0^{\frac{1}{1-r}}$ vs. $r$ plane.  In the non-interacting limit, we have shown in section 3 that for $r<1$ the ground state is SC, while for $r>1$ it is LM.  From finite-$U$ NRG studies [12,13] by contrast, it is known that the ground state is exclusively LM for $r>\frac{1}{2}$.  In consequence, for $r>1$, second order perturbation theory in $U$ about the non-interacting limit as considered in section 5 appears able to capture at least the intial evolution of the LM state upon increasing $U$ from zero, yielding in particular the characteristic low-frequency spectral signature of the LM phase found in NRG studies [12], viz $D(\omega)\sim |\omega|^r$.  For $\frac{1}{2}<r<1$ by contrast, the $U=0$ and $U>0$ ground states are fundamentally distinct, and we believe a finite-order perturbative treatment about the non-interacting limit to be simply inapplicable.

Perhaps the most subtle regime is $r<\frac{1}{2}$, encompassing as a particular case the `normal' Anderson model, $r=0$.  In section 4, general conditions were established for a SC state to be realized for $U>0$.  Physically, these amount to interactions having no influence in renormalizing the lowest-frequency asymptotic behaviour of the impurity spectrum $D(\omega)$, whose behaviour as $\omega \rightarrow 0$ is that of the non-interacting limit, viz $D(\omega) \sim |\omega|^{-r}$---producing the characteristic spectral signature of the SC state observed in NRG studies [12].  Significantly, one deduces in consequence a pinning condition (equation (4.7)) upon $A(\omega) = |\omega|^rD(\omega)$, whereby $A(\omega = 0)$ at the Fermi level is pinned at its non-interacting value for all $U$ and $r$ where a SC state obtains, and which represents a natural generalization of the familiar pinning condition $D(\omega=0)=1/\pi\Delta_0\  \forall\  U$ characteristic of the $r = 0$ Anderson model.  At the level of second order perturbation theory in $U$ (section 5.2), such a state of affairs is realized in practice for $r <\frac{1}{2}$, where the characteristic spectra $A(\omega)$ (figure 5) are strikingly  reminiscent of their counterparts for $r=0$ (figure 6), with evolving Hubbard satellites upon increasing $U$ and the emergence of a low-frequency Kondo scale reflected in the progressively narrowing width of the generalized Abrikosov-Suhl resonance in $A(\omega)$.

\restylefloat{figure}
\begin{figure}[H] 
\centering 
\epsfig{file =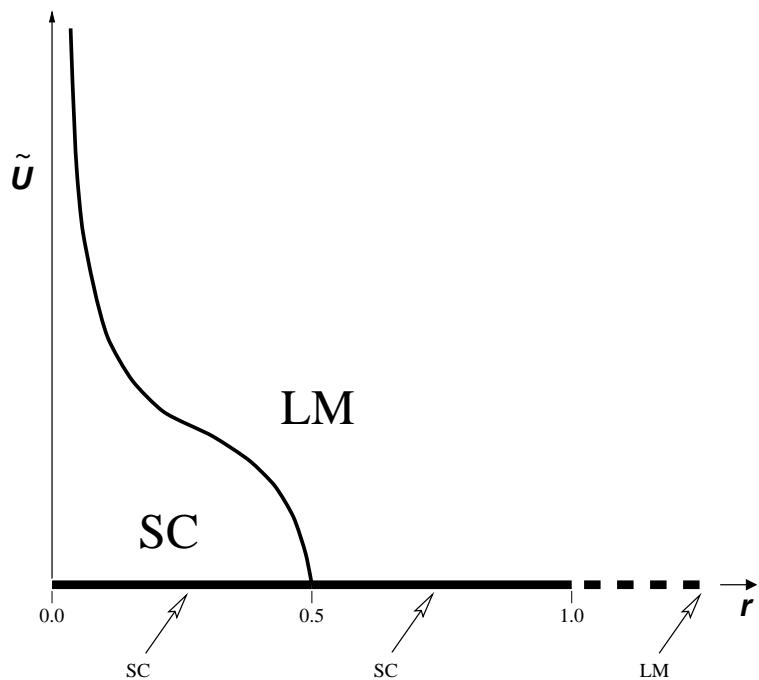,width=10cm} 
\caption{Schematic phase boundaries between SC and LM states in the $\tilde{U}=U/\Delta_0^{\frac{1}{1-r}}$ versus $r$ plane; details in text.}
\end{figure}

Figure 7 also shows clearly the limitations of finite order perturbation theory in the interaction strength.  For $r<\frac{1}{2}$, NRG studies yield both SC and LM phases [12,13], the critical line $\tilde{U}_C(r)$ diverging as $r \rightarrow 0$, reflecting the fact that the $r=0$ Anderson model is a Fermi liquid for all finite $U$; and vanishing as $r \rightarrow \frac{1}{2}$ (although whether it does so continuously as $r \rightarrow \frac{1}{2}-$ we regard as not wholly settled).  It is in part this feature---the existence of a transition between a LM state and a SC (or generalized Fermi liquid) state---that renders the problem of generic interest.  But low order perturbation theory, capable though it is of describing the initial evolution of the SC state upon increasing $U$ from zero, will naturally produce a SC state for all $U$ and cannot therefore delineate its boundary.  To deal with this central issue, as well as to describe successfully the regime $\frac{1}{2}<r<1$, one requires an inherently non-perturbative theory that is capable of capturing both LM and generalized Fermi liquid phases, and hence the transition between them.  For this, we believe a rather radical departure from conventional theoretical approaches is required, and will turn to one such in a subsequent paper [14].

\ack
We are grateful to R Bulla and T Pruschke for stimulating discussions.  Financial support from the EPSRC and British Council is also gratefully acknowledged.

\end{document}